\documentclass[twocolumn,prl,aps]{revtex4}
\usepackage{graphicx}
\begin{document}
\renewcommand{\thefootnote}{\fnsymbol{footnote}}
\sloppy

\newcommand \be{\begin{equation}}
\newcommand \bea{\begin{eqnarray} \nonumber }
\newcommand \ee{\end{equation}}
\newcommand \eea{\end{eqnarray}}
\newcommand{\rar}{\rightarrow}
\newcommand{\eq}{equation}
\newcommand{\eqs}{earthquakes}
\newcommand{\rp}{\right)}
\newcommand{\lp}{\left(}

\title{On the Occurrence of Finite-Time-Singularities in
Epidemic Models of Rupture, Earthquakes and Starquakes}
\author{D. Sornette$^{1,2}$ and A. Helmstetter$^{3}$}
\address{$^1$ Department of Earth and Space Sciences and Institute of
Geophysics and Planetary Physics University of California, Los Angeles,
California 90095-1567}
\address{$^2$ Laboratoire de Physique de la Mati\`{e}re
Condens\'{e}e, CNRS UMR 6622 
Universit\'{e} de Nice-Sophia Antipolis, Parc Valrose, 06108 Nice,
France}
\address{$^3$ Laboratoire de G{\'e}ophysique Interne et Tectonophysique,
Observatoire de Grenoble, Universit\'e Joseph Fourier, France}
\begin{abstract}
We present a new kind of
critical stochastic finite-time-singularity, relying on the interplay
between long-memory and extreme fluctuations.
We illustrate it on
the well-established epidemic-type aftershock (ETAS) model for aftershocks,
based solely on the most solidly documented stylized facts of
seismicity (clustering in space and in time and power law Gutenberg-Richter
distribution of earthquake energies). This theory accounts for the
main observations
(power law acceleration and discrete scale invariant structure) of critical
rupture of heterogeneous materials, of the largest
sequence of starquakes ever attributed to a neutron star as well as of
earthquake sequences.
\end{abstract}

\maketitle

\vskip 0.5cm

A large portion of the current work on rupture and earthquake 
prediction is based on the
search for precursors to large events in the seismicity itself. Observations of
the acceleration of seismic moment leading up to large events and ``stress
shadows'' following them have been interpreted as evidence that seismic cycles
represent the approach to and retreat from a critical state of a fault network
\cite{Sorsammis}. This ``critical state'' concept is fundamentally
different from
the long-time view of the crust as evolving spontaneously in a statistically
stationary critical state, called self-organized criticality (SOC)
\cite{BakTangSorSor}. In the SOC view, all events belong to the same global
population and participate in shaping the self-organized critical state. Large
earthquakes are inherently unpredictable because a  big  earthquake
is simply a small
earthquake that did not stop. By
contrast, in the critical point view, a great earthquake plays a special
role and signals the end of a cycle on its fault network. The dynamical
organization is not statistically stationary but evolves as the great 
earthquake
becomes more probable. Predictability might then become possible by monitoring
the approach of the fault network towards the critical state. This hypothesis
first proposed in \cite{Sorsammis} is
the theoretical induction of a series of observations of accelerated seismicity
\cite{SykesJaum1,BufeVarnes} which has been later strengthened by several other
observations \cite{HarrisSimpson,Bowmanetal,Zoller,Yin}. Theoretical
support has also come from simple computer models of critical rupture
\cite{critrupd} and experiments of material rupture \cite{expcrit},
cellular automata, with \cite{Anghel}
and without \cite{Huang} long-range interaction, and from granular simulators
\cite{Mora}. Models of regional seismicity with more faithful fault
geometry have
been developed that also show accelerating seismicity before large model events
\cite{Heibenkin,BK,benzion}.

There are at least five different mechanisms that are known to lead to critical
accelerated seismicity of the form
\be
N(t) \propto 1/(t_c -t)^m
\label{powerlawrate}
\ee
ending at the critical time $t_c$, where $N(t)$ is the seismicity
rate (or acoustic emission rate for material rupture).
Such finite-time-singularities are quite common and have
been found in many well-established models of natural systems, either
at special
points in space such as in the Euler equations of inviscid fluids, in vortex
collapse of systems of point vortices, in the equations of General Relativity
coupled to a mass field leading to the formation of black holes, in models of
micro-organisms aggregating to form fruiting bodies, or in the more prosaic
rotating coin (Euler's disk). They all involve
some kind of positive feedback, which in the rupture context
can be the following (see \cite{SamsorPNAS} for a review):
sub-critical crack growth \cite{DasScholz},
geometrical feedback in creep rupture \cite{Krajcinovic},
feedback of damage on the elastic coefficients
with strain dependent damage rate \cite{benzion},
feedback in a percolation model of regional seismicity \cite{SamsorPNAS},
feedback in a stress-shadow model for regional seismicity \cite{BK,SamsorPNAS}.

While these mechanisms are plausible, their relevance to the earth
crust remains
unproven. Here, we present a novel mechanism leading to a new kind of
critical stochastic finite-time-singularity in the seismicity rate, using
the well-established epidemic-type aftershock sequence (ETAS) model
  for aftershocks,  introduced by \cite{KK,Ogata1},
based solely on the most solidly documented stylized facts of
seismicity mentioned above. The adjective ``stochastic'' emphasizes
the fact that the critical time $t_c$ is determined in large part
by the specific sets of innovations of the random process. We show that,
in a finite domain of its parameter space, the rate of seimic activity
in the ETAS model diverges in finite time
according to (\ref{powerlawrate}). The underlying mechanism relies on
large deviations
occuring in an explosive branching process.
One of the advantage of this discovery is to
be able to account for the observations of accelerated seismicity
and acoustic emission in material failure, without invoking
any new ingredient other than those already well-established
empirically.
We apply this insight to quantify the longest available starquake
sequence of a neutron star soft $\gamma$-ray repeaters.

We shall use the example of earthquakes but the model applies similarly to
microcracking in materials.
The ETAS model is a generalization of the modified Omori law, in that it takes
into account the secondary aftershock sequences triggered by all events.
The modified Omori's law states that
the occurrence rate of the direct aftershock-daughters from
an earthquake decreases with the time from the mainshock
according to the ``bare propagator'' $K/{(t+c)}^p$.
In the ETAS model, all earthquakes are simultaneously
mainshocks, aftershocks and possibly foreshocks.
Contrary to the usual definition of aftershocks, the ETAS model
does not impose an aftershock
to have an energy smaller than the mainshock.
This way, the same law describes both foreshocks, aftershocks and mainshocks.
An observed ``aftershock''
sequence of a given earthquake (starting the clock)
is the result of the activity of all events
triggering events triggering themselves other events, and so on,
taken together.
The corresponding seismicity rate (the ``dressed propagator''), which
is given by the superposition of the aftershock
sequences of all events, is the quantity we derive here.

Each earthquake (the ``mother'') of energy
$E_i \geq E_0$ occurring at time $t_i$
gives birth to other events (``daughters'') of any possible
energy, chosen with the Gutenberg-Richter
density distribution $P(E) = \mu/(E/E_0)^{1+\mu}$ with exponent $\mu 
\simeq 2/3$,
at a later time between $t$ and $t+dt$  at the rate
\be
\phi_{E_i}(t-t_i) = \rho(E_i)~\Psi(t-t_i)~.
\label{first}
\ee
$\rho(E_i) =K ~(E_i/E_0)^a$ gives the number of daughters born
from a mother with energy $E_i$, with the same exponent $a$ for all
earthquakes.
This term accounts for the fact that large mothers have many more
daughters than small mothers because the larger spatial extension of 
their rupture
triggers a larger domain. $E_0$ is a lower bound energy below which no daughter
is triggered. $\Psi(t-t_i) = {\theta~c^{\theta}  \over (t-t_i+c)^{1+\theta}}$
is the normalized waiting time distribution (local Omori's law or 
``bare propagator'')
giving the rate
of daughters born a time $t-t_i$ after the mother.

The ETAS model is fundamentally a ``branching'' model \cite{Branching}
with no ``loops'', i.e.,
each event has a unique ``mother-mainshock'' and not several. This
``mean-field'' or random phase approximation allows us to simplify the analysis
while still keeping the essential physics in a qualitative way. The problem
is to calculate the ``dressed'' or ``renormalized'' propagator (rate of seismic
activity) that includes the whole cascade of secondary sequences \cite{sorsor}.
The key parameter is the average number $n$ (or
``branching ratio'') of daughter-earthquakes
created per mother-event, summed over all possible energies.
$n$ is equal to the integral of $\phi_{E_i}(t-t_i)$
over all times after $t_i$ and over all energies $E_i \geq E_0$.
This integral converges to a finite value $n < \infty$ for $\theta>0$
(local Omori's law decay
faster than $1/t$) and for $a<\mu$ (not too large a growth of the number
of daughters
as a function of the energy of the mother).
The resulting average rate $N(t)$  of seismicity is the
solution of the Master equation \cite{helmsor}
\be
N(t) = \int_0^t d\tau~N(\tau)  \int_{E_0}^{E_{\rm max}(t)} dE'~
P(E')~\phi_{E'}(t-\tau)
\label{thirdter}
\ee
giving the number $N(t) dt $ of events occurring between $t$ and $t+dt$ of any
possible energy. We have made explicit the upper bound $E_{\rm max}(t)$ equal
to the typical maximum earthquake energy sampled up to time $t$.
For $a < \mu$, this upper bound has no impact on the results and can be 
replaced by $+\infty$ \cite{helmsor}.
There may be a source term $S(t)$ to add to the r.h.s. of
(\ref{thirdter}), corresponding to
either a constant background seismicity or to a large triggering earthquake.
In this last case, the rate $N(t)$ solution of (\ref{thirdter}) is the
``dressed'' propagator giving the renormalized Omori's law.
A rich behavior, which has
been fully classified by a complete analytical treatment \cite{helmsor},
has been found: sub-criticality $n<1$ \cite{sorsor}
and super-criticality $n > 1$ \cite{helmsor}, where $n$ depends on the
  control parameters $\mu$, $a$, $\theta$, $K$ and $c$.
With a single value of the exponent $1+\theta$ of
the ``bare'' propagator $\Psi(t) \sim 1/t^{1+\theta}$,
we obtain a continuum of apparent exponents for the global rate of aftershocks
\cite{helmsor}
which may account for the observed variability of Omori's exponent 
$p$ around $p=1$
reported by many workers.

Here, we explore the regime $a \geq \mu$, for which $n$ is infinite.
This signals the impact of large earthquake energies, suggesting
the relevance of the upper bound $E_{\rm max}(t)$ in (\ref{thirdter}).
This case is actually observed
in real seismicity by \cite{Drakatos} who obtained $a >\mu$
for some aftershock sequences in Greece, and by \cite{guoogata}
who found $a >\mu$ for 13 out of 34 aftershock sequences in Japan.
This case $ a >\mu$ also characterizes
  the seismic activity preceding the 1984
$M=6.8$ Nagano Prefecture earthquake \cite{Ogata2}. After the
mainshock, the seismicity returned in the sub-critical regime
  $\theta>0$, $a<\mu$ and $n<1$.

This case $a\geq \mu$ is similar to that found
underlying various situations of anomalous transport \cite{BG}: in this
regime of large fluctuations, 
the integral over earthquake energies is dominated
by the upper bound. The maximum
energy $E_{\rm max}(t)$ sampled by $N(t) \Delta t$ earthquakes
is given by the standard condition $N(t) \Delta t \int_{E_0}^{E_{\rm max}(t)} dE' ~P(E')
\approx 1$.
This yields the robust median estimate $E_{\rm max}(t) \sim
[N(t) \Delta t]^{1/\mu}$.
Actually, $E_{\rm max}(t)$ is itself distributed according to the
Gutenberg-Richter distribution and thus exhibits large fluctuations
from realization to realization, as we can see in Fig.~1. Putting this
estimation of $E_{\rm max}(t)$ in (\ref{thirdter}), we get
\be
N(t) \propto \int_0^t d\tau~ {N(\tau)  \over (t-\tau +c)^{1+\theta}}~
[N(\tau) \Delta \tau]^{(a -\mu)/\mu}~.   \label{jjsla}
\ee
Let us note the appearance of the new term $[N(\tau) \Delta \tau]^{(a -\mu)/\mu}$
resulting from the contribution of the upper bound in the integral
$\int dE' P(E')$. This term replaces the constant found for the case $a < \mu$.
Equation (\ref{jjsla}) shows that the
exploration of larger and
larger events in the tail in the Gutenberg-Richer distribution
transforms the {\it linear}
Master equation (\ref{thirdter}) into a {\it non-linear} equation:
the non-linearity expresses
a positive feedback according to which the larger is the rate $N(t)$
of seismicity,
the larger is the maximum sampled earthquake, and the larger is the
number of daughters
resulting from these extreme events. This process self-amplifies and
leads to the
announced finite-time singularity (\ref{powerlawrate}). However, to
complete the derivation, we
need to determine the yet unspecified time increment $\Delta \tau$. If
$N(\tau)$ obeys (\ref{powerlawrate}), $\Delta \tau$ is not a constant
that can be
factorized away: it is determined by the condition that, over $\Delta \tau$,
$N(\tau)$ does not change ``significantly'' in the interval 
$[\tau, \tau+\Delta \tau]$, i.e., no more than by a
constant factor.
Using the assumed power law solution (\ref{powerlawrate}), this gives
$\Delta \tau \propto t_c - \tau$. Using this and
inserting (\ref{powerlawrate}) into (\ref{jjsla}), we get,
\bea
m &=& {a/\mu \over (a/\mu)-1}~,~~~~~~~~~~~~~~~~~~t_c-t \leq c  \nonumber    \\
m &=& {(a/\mu) - 1 -\theta ~H(-\theta) \over (a/\mu)-1}~, ~~~ t_c-t \gg c~,
\label{predim}
\eea
where $H$ is the Heaviside function.
Note that (\ref{predim}) predicts an exponent $m>1$ which is
independent of $\theta$
close to the critical time $t_c$. This is due to the fact that
the time decay of the Omori's kernel is not felt for $t_c-t \leq c$,
where $c$ acts
as an ultraviolet cut-off. It is also interesting to find that $m=1$
independently of
$a$ and $\theta$ in the regime $\theta >0$ (with of course $a>\mu$) for which
Omori's kernel $\sim 1/t^{1+\theta}$ decays sufficiently fast
at long times that the predominant
contributions to the present seismic rate come from
events in the immediate past of the present time of observation. In 
constrast, the case
$\theta<0$ is
analogous to the anomalous long-time memory regime \cite{BG}
which keeps for ever the impact of past events on future rates.

This prediction, based on the careful analysis
of the integral in (\ref{jjsla}), has been verified by direct
numerical evaluation
of the equation (\ref{jjsla}). We have also checked that numerical
Monte Carlo simulations of the
ETAS model generates catalogs of events following this prediction, in
an ensemble or median sense. Figure 1 shows the cumulative number
${\cal N}(t) = \int_0^t d\tau ~N(\tau)$
of events for a typical realization of the ETAS model and compares
it with $E_{\rm max}(t)$ to illustrate that ${\cal N}(t)$
is mostly controlled by
the sampling of $E_{\rm max}(t)$, as discussed in the derivation of 
expression (\ref{jjsla}) leading to the
finite-time-singularity (\ref{powerlawrate}).
 For the value $\mu=1$ chosen here, $E_{\rm max}(t)$ follows
the same power law as the cumulative number, as observed.
The dashed line
is the power law prediction (\ref{powerlawrate}) with (\ref{predim}) 
for $a/\mu = 1.5$
and $\theta =-0.2$ with slope $m-1=0.4$.
  We have also generated 500 such catalogs and report in the inset the
distribution $d(m)$ of exponents $m$ obtained by a best fit of ${\cal
N}(t)$ for each of the 500 catalogs to
a power law $1/(t_c-t)^{m-1}$. The median of $d(m)$ is exactly equal to the
prediction shown by the vertical thin line while the mode is very close to it.
Note however a rather large dispersion which is expected from the highly
intermittent dynamics characteristic of this extreme-dominated dynamics.
We now report a few comparisons between
the prediction (\ref{predim}) and the median value of the exponent $m$
obtained from 500 simulations for the following parameters:
$\theta=-0.2$, $a=1.7, \mu=1$, predicted $m=1.29$, median $m=1.37$;
$\theta=-0.2$, $a=1.3, \mu=1$, predicted $m=1.67$, median $m=1.61$;
$\theta=-0.1$, $a=1.5, \mu=1$, predicted $m=1.20$, median $m=1.29$;
$\theta=-0.3$, $a=1.5, \mu=1$, predicted $m=1.60$, median $m=1.62$.
For $a>1.8\mu$, the fluctuations are so large that a reliable 
determination of the median becomes questionable from a sample of 500 realizations
and many more would be needed.


Figure 1 shows that the power law singularities are
decorated by quite strong steps or oscillations, approximately
equidistant in the
variable $\ln (t_c-t)$. This log-periodicity has been previously proposed as
a possibly important signature of rupture and earthquake sequences
approaching a critical point \cite{Sorsammis,expcrit}.
Here, we present a simple novel mechanism for this observation, based
on a refinement of the previous
argument leading to $E_{\rm max}(t) \sim [N(t) \Delta t]^{1/\mu}$. Indeed,
the most probable value for the energy $E_n$ of the
$n$-th largest earthquake ranked from the largest $E_1=E_{\rm max}$
to the smallest one is given
by $E_n(t) = \{[{\cal N}(t) \mu + 1]/[n \mu + 1]\}^{1/\mu}$ \cite{rank},
where ${\cal N}(t) = \int_0^t N(t') dt'$.
Let us assume that the last new record was broken at time $t_1$ leading to
$E_1(t_1) = \{[{\cal N}(t_1) \mu + 1]/[\mu + 1]\}^{1/\mu}$. The next
record will occur
at a time $t_2>t_1$ whose typical value is
such that $E_2(t_2) = E_1(t_1)$ (the
last record $E_1(t_1)$ becomes the second largest event $E_2(t_2)$
when a new record $E_1(t_2)$ occurs).
For large ${\cal N}(t)$, this gives ${{\cal N}(t_2) \over {\cal
N}(t_1)} = (2 \mu +1)/(\mu +1)$.
The prefered scaling ratio of the average log-periodicity is
$\lambda \equiv (t_c-t_1)/(t_c-t_2) = [(2\mu + 1)/(\mu +1)]^{1/(m-1)}$.
For $a/\mu=1.5, \theta=-0.2, m=1.4$ corresponding to figure 1,
we obtain $\lambda \approx 2.3$, which is compatible with the data.

The prediction (\ref{predim}) rationalizes the ``inverse'' Omori's law close to
$1/(t_c-t)$ that has been documented for earthquake forshocks \cite{KKpred}.
The prediction (\ref{predim}) as well as the log-periodicity offers a
general framework to rationalize
several previous experimental reports of precursory acoustic emissions rates
prior to global failures \cite{expcrit}. In this case, the
energy release rate $e(t)$ is found to follow a power law finite-time
singularity. According to our theory, $e(t) \propto N(t) E_{\rm max}(t) \propto
1/(t_c-t)^{m+(m-1)/\mu}$.

Finally, we also show that this could explain starquakes catalogs.
Starquakes are assumed to be
ruptures of a super-dense $1$-km thick crust made of heavy nuclei
stressed by super-strong
stellar magnetic field. They are observed through the associated
flashes of soft $\gamma$-rays radiated during the rupture.
Starquakes exhibit all the main stylized facts of
their earthly siblings
\cite{Kosso}. The thick line in figure 1 shows the cumulative number of
starquakes of the SGR1806-20 sequence, which is the longest sequence
  (of 111 events) ever attributed to the same neutron star, as a function of the
logarithm of the time $t_c-t$ to failure.  The
starquake data is compatible with $\mu=1$ \cite{Kosso}, $a = 1.5$ and
$\theta =-0.2$, leading to $m = 1.4$.

We are grateful to V. Keilis-Borok and V. Kossobokov for sharing the starquake
data with us and W.-X. Zhou for discussions and help in a preliminary
analysis of the data.

\begin{figure}
\begin{center}
\includegraphics[width=8cm]{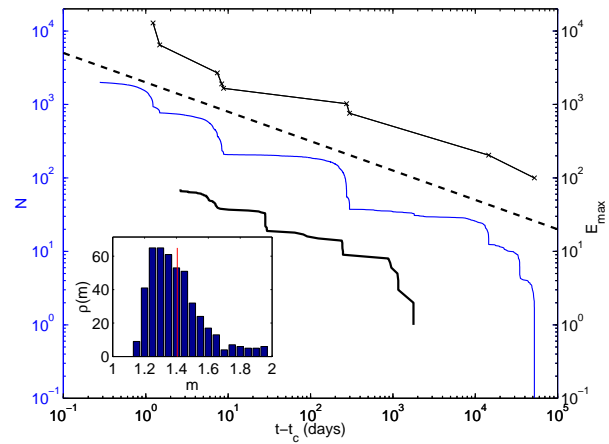}
\end{center}
\caption{Cumulative number of events (scale on the left) as a function of the time from
the critical
point $t_c$ for the starquake sequence (solid black line) and one typical
simulation of the ETAS model (solid thin line) generated with
$\theta=-0.2$, $a/\mu=1.5$ and $c=0.001$. For the starquakes, $t_c$ is the time of
the strongest observed starquake in the sequence.
The dashed line shows the theoretical exponent $m-1=0.4$ (\ref{predim})
for $t_c-t>c$. The crosses $\times$ joined by straight segments give the
time evolution of $E_{\rm max}(t)$ (scale on the right). 
The inset gives the distribution of exponent measured for 500 numerical
simulations. The median (vertical line) of the distribution of $m$-values is equal to the
theoretical exponent $m=1.4$ (formula (\ref{predim})).
}
\label{fig1}
\end{figure}

\end{document}